\documentclass[11pt]{amsart}
\usepackage[numbers]{natbib}
\renewcommand{\baselinestretch}{1.2}
\begin{document}

\title{A Simple Proof of the F{\'a}ry-Wagner  Theorem}

\author{David R. Wood}
\address{Departament de Matem{\`a}tica Aplicada II, Universitat Polit{\`e}cnica de Catalunya, Barcelona, Spain}
\thanks{Supported by grant MEC SB2003-0270. Partially completed at McGill University, Montr\'eal, Canada}
\email{david.wood@upc.edu}
\date{\today}
\maketitle

The purpose of this note is to give a simple proof of the following fundamental result independently due to \citet{Fary48} and \citet{Wagner36}. A \emph{plane graph} is a simple graph embedded in the plane without edge crossings. Combinatorially speaking, there is a circular ordering of the edges incident to each vertex, and a nominated outerface.

\medskip\noindent
\textbf{Theorem}. Every plane graph has a drawing in which every edge is straight.

\medskip\noindent
\emph{Proof}. A \emph{triangulation} is a plane graph in which every face is bounded by three edges. Edges can be added to a plane graph to obtain a plane triangulation. Thus it suffices to prove the theorem for plane triangulations $G$. We proceed by induction on $|V(G)|$. The base case with $|V(G)|=3$ is trivial. Now suppose that $|V(G)|\geq4$. A \emph{separating triangle} of $G$ is a $3$-cycle that contains a vertex in its interior and in its exterior. If $G$ has no separating triangles, then let $vw$ be any edge of $G$. Otherwise, let $vw$ be an edge incident to a vertex that is in the interior of an innermost separating triangle of $G$. Now $vw$ is on the boundary of two faces, say $vwp$ and $vwq$. Since $vw$ is not in a separating triangle, $p$ and $q$ are the only common neighbours of $v$ and $w$. Let $(vp,vw,vq,vx_1,vx_2,\dots,vx_k)$ and $(wq,wv,wp,wy_1,wy_2,\dots,wy_\ell)$ be the clockwise ordering of the edges incident to $v$ and $w$ respectively\footnote{In fact, for every vertex $v$ there is an edge incident to $v$ whose endpoints have at most two common neighbours. This is because the neighbourhood of $v$ has no $K_4$-minor (it is even outerplanar), and every graph with no $K_4$-minor has a vertex of degree at most two.}.

Let $G'$ be the plane triangulation obtained from $G$ by contracting the edge $vw$ into a single vertex $s$. Replace the pairs of parallel edges $\{vp,wp\}$ and $\{vq,wq\}$ in $G$ by edges $sp$ and $sq$ in $G'$. The clockwise ordering of the edges of $G'$ incident to $s$ is $(sp,sy_1,sy_2,\dots,sy_\ell,sq,sx_1,sx_2,\dots,sx_k)$. By induction, $G'$ has a drawing in which every edge is straight (and the circular ordering of the edges incident to $s$ are preserved). For all $\epsilon>0$, let $C_\epsilon(s)$ denote the circle of radius $\epsilon$ centred at $s$. For each neighbour $t$ of $s$ in $G'$, let $R_\epsilon(t)$ denote the region consisting of the union of all open segments between $t$ and a point in $C_\epsilon(s)$. There is an $\epsilon>0$ such that all neighbours $t$ of $s$ are in the exterior of $C_\epsilon(s)$ and the only edges of $G'$ that intersect $R_\epsilon(t)$ are incident to $s$.

There is a line $L$ through $s$ with $p$ on one side of $L$ and $q$ on the other side, as otherwise the edges $sp$ and $sq$ would overlap. Now $sp$ and $sq$ break $C_\epsilon(s)$ into two arcs, one that intersects the edges $\{sx_i:1\leq i\leq k\}$, and one that intersects the edges $\{sy_j:1\leq j\leq \ell\}$. The set $L\cap C_\epsilon(s)$ consists of two points. Position $v$ and $w$ at these two points, with $v$ on the side of $C_\epsilon(s)$ that intersects the edges $\{sx_i:1\leq i\leq k\}$, and with $w$ on the other side. Delete $s$ and its incident edges. Draw the edges of $G$ incident to $v$ or $w$ straight. Thus $vw$ is contained in $L$. Since $p$ and $q$ are on different sides of $L$, the edges incident to $v$ or $w$ do not cross. By the choice of $\epsilon$, edges incident to $v$ or $w$ do not cross other edges of $G$. Thus we obtain the desired drawing of $G$.\qed


\def\soft#1{\leavevmode\setbox0=\hbox{h}\dimen7=\ht0\advance \dimen7
  by-1ex\relax\if t#1\relax\rlap{\raise.6\dimen7
  \hbox{\kern.3ex\char'47}}#1\relax\else\if T#1\relax
  \rlap{\raise.5\dimen7\hbox{\kern1.3ex\char'47}}#1\relax \else\if
  d#1\relax\rlap{\raise.5\dimen7\hbox{\kern.9ex \char'47}}#1\relax\else\if
  D#1\relax\rlap{\raise.5\dimen7 \hbox{\kern1.4ex\char'47}}#1\relax\else\if
  l#1\relax \rlap{\raise.5\dimen7\hbox{\kern.4ex\char'47}}#1\relax \else\if
  L#1\relax\rlap{\raise.5\dimen7\hbox{\kern.7ex
  \char'47}}#1\relax\else\message{accent \string\soft \space #1 not
  defined!}#1\relax\fi\fi\fi\fi\fi\fi} \def\cprime{$'$}

\end{document}